\documentclass[fleqn,usenatbib]{mnras}

\usepackage{newtxtext,newtxmath}

\usepackage[T1]{fontenc}
\usepackage{ae,aecompl}

\usepackage{soul}

\usepackage[utf8]{inputenc}
\usepackage{graphicx}	
\usepackage{amsmath}	
\usepackage{amssymb}	
\usepackage{pdflscape}	

\newcommand*\samethanks[1][\value{footnote}]{\footnotemark[#1]}

\title[Formation of SMBH seeds via collisions: importance of mass loss]{Formation of SMBH seeds in Population III star clusters through collisions :  the importance of mass loss}

\author[P. J. Alister Seguel et al.]{P. J. Alister Seguel ,$^{1}$\thanks{E-mail: patalister@udec.cl (PJAS); dschleicher@astro-udec.cl (DRGS)}
D. R. G. Schleicher,$^{1}$\samethanks[1]
 T. C. N. Boekholt,$^{2,3}$ \newauthor 
M. Fellhauer$^{1}$   \&
R. S. Klessen$^{4,5}$
\\
$^{1}$Departamento de Astronomía, Facultad de Ciencias Físicas y Matemáticas, Universidad de Concepción, Av. Esteban Iturra s/n \\ Barrio  Universitario, Casilla 160-C, Concepción, Chile\\
$^{2}$Instituto de Telecomunica\c{c}\~oes, Campus Universit\'ario de Santiago, 3810-193, Aveiro, Portugal\\
{$^{3}$Department of Physics, University of Coimbra, 3004-516, Coimbra, Portugal } \\
$^{4}$Universität Heidelberg, Zentrum für Astronomie, Institut für Theoretische Astrophysik, Albert-Ueberle-Str. 2, D-69120 Heidelberg,\\ Germany\\
$^{5}$ Universität Heidelberg, Interdisziplinäres Zentrum für Wissenschaftliches Rechnen, Im Neuenheimer Feld 205, D-69120 Heidelberg,\\ Germany
}

\date{Accepted XXX. Received YYY; in original form ZZZ}

\pubyear{2019}

\begin{document}
\label{firstpage}
\pagerange{\pageref{firstpage}--\pageref{lastpage}}
\maketitle

\begin{abstract}
Runaway collisions in dense clusters may lead to the formation of supermassive black hole (SMBH) seeds, and this process can be further enhanced by accretion, as recent models of SMBH seed formation in Population III star clusters have shown. This may explain the presence of supermassive black holes already at high redshift, $z>6$. However, in this context, mass loss during collisions was not considered and could play an important role for the formation of the SMBH seed. Here, we study the effect of mass loss, due to collisions of protostars, in the formation and evolution of a massive object in a dense primordial cluster. We  consider both constant mass loss fractions as well as analytic models based on the stellar structure of the collision components. Our calculations indicate that mass loss can significantly affect the final mass of the possible SMBH seed. Considering a constant mass loss of 5\% for every collision, we can lose between 60-80\% of the total mass that is obtained if mass loss were not considered. Using instead analytical prescriptions for mass loss, the mass of the final object is reduced by 15-40\%, depending on the accretion model for the cluster we study. Altogether, we obtain masses of the order of $10^4M_{\odot}$, which are still massive enough to be SMBH seeds.
\end{abstract}




\section{Introduction}

A great number of supermassive black holes (SMBH) has already been
detected at redshifts higher than 6 ($z = 7.085$ in \citealt{mortlock}, $z=7.54$ in \citealt{banados}), and the number is 
still continuously increasing. These high redshifts roughly correspond to the first billion years of evolution of the Universe.
A significant challenge to our cosmological model is explaining the existence of such massive objects at early times; if we consider the Eddington accretion rate, which is the maximum rate at which a black hole can accrete gas  in spherical symmetry, initial seed masses of order $10^4M_{\odot}$ are required in order to reach final masses of $10^9M_{\odot}$, when realistic spin parameters and accretion disk models are taken into account \citep{shapiro}. The
only solutions are very massive seeds or extended periods
of super-Eddington accretion, which can persist in non-spherically
symmetric circumstances where the Eddington rate can be breached, or possible combinations
between the two in the early stages of the massive
black hole evolution \citep{mayer}.

The three main pathways for the formation of supermassive
black holes are:  stellar remnants in the context of massive Population III (in short Pop. III) stars, the collapse of a protogalactic gas cloud into a massive black hole or a massive star, that later collapses into a black hole, and seed black holes forming in dense
stellar clusters via dynamical processes  \citep{rees,Woods_2018}.
One of the most promising explanations for massive seeds is the direct collapse model, as it can potentially produce the most massive seeds.
In order to form massive black holes in this way, numerical simulations have shown that cooling needs to be suppressed, which can be achieved when a strong radiation background  photodissociates the molecular hydrogen, therefore preventing strong fragmentation of the gas cloud \citep{brom,wise,Latif_2013}.

 The strength of the radiation background required to keep the
gas atomic is parametrized via $J_{21}$, which describes the
strength of the radiation background at 13.6 eV. A value of $J_{21}=1$ corresponds to a radiative background of $10^{-21} $erg s$^{-1}$ cm$^{-1}$ Hz$^{-1}$ per steradian. First investigations using numerical simulations suggested a critical value of $J_{21} \sim$ 100 to prevent the
formation of molecular hydrogen \citep{shang2010}, later investigations which considered updated chemical
networks and more realistic models for the radiation background \citep[see e.g.][]{sugimura,agarwal15} have given much larger critical values, of the order $10^5$
when applied in cosmological simulations \citep{latif14,latif15}. One of the biggest problems for the direct collapse scenario is the need
of a large value of $J_{21}$ \citep{dijkstra}.

Furthermore, metals and dust grains can also be the cause of fragmentation events. For metal line cooling, a metallicity of
$10^{-3}$ Z$_\odot$ can already trigger fragmentation within cosmological simulations \citep{bovino14}, and if dust cooling is considered, the metallicities required are considerably lower, of the order of $10^{-5}$ Z$_\odot$  \citep{schneider,dopcke,bovino16}. 
The need of both having very strong radiation backgrounds,
while keeping the gas metal free, leads to a strong need of
fine-tuning, which at best can be satisfied under very rare
conditions \citep[eg,][]{agarwal17}.
Recently, \cite{suazo2019role} studied the formation of SMBH seeds in this context, forming a single massive object of $\sim 10^5 M_{\odot}$  when the UV background of $J_{21}$ is set to 10000, while when considering a UV background of $J_{21} = 10$, there is
fragmentation and the formation of various less massive seeds. These fragments had masses of $10^3-10^4 M_{\odot}$, and even though less massive, they were still prone to merge into a more massive object.

The alternative pathway of early black hole formation in stellar clusters via dynamical processes has been studied to a smaller extent. \cite{devecchi09}
and \cite{devecchi12} developed analytical models which predict black hole masses of
$100 - 1000 M_{\odot}$ forming in the first stellar clusters. \cite{katz} modeled the formation of a dense stellar cluster in a cosmological simulation, and showed the subsequent
formation of a $\sim1000 M_{\odot}$ black hole via $N$-body simulations. Likewise, \cite{sakurai} combined cosmological and $N$-body simulations which led to the formation
of black holes in the first stellar clusters with $400 - 1900$ $M_{\odot}$. \cite{reinoso} studied collisions in massive Pop. III
clusters, showing there could be resulting black hole masses of up to $600 M_{\odot}$.

\cite{tjarda} were the first to explore
the formation of massive black hole seeds from a dense stellar cluster, taking into account the interaction between gas-dynamical and stellar-dynamical processes, considering also the accretion of the protostars and consequently enhanced radii.
Since the initially low mass Pop. III protostars would gain mass
by accreting from the gas reservoir, and as the accretion rate
might vary with cluster environment and cluster evolution, they
defined 6 different accretion models, in which accretion depends on the gas availability and position of the protostar. The models are further described in Section \ref{sec:2} and summarized in table \ref{tab:tjar}. They concluded that accretion-induced
collisions in dense Pop. III protostellar systems, in the presence of a sufficiently large gas reservoir, are a viable
mechanism for explaining the formation of the first massive
black hole seeds. Stellar collisions in primordial star clusters can give
rise to the formation of massive objects of $10^4 - 10^5 M_{\odot}$ for all the models considered.

This investigation warranted follow up studies, to improve on the realism of the
implementation and to include additional physics previously not considered. One important aspect not taken into account was the mass that might get lost whenever two protostars collide, which is the main focus of the study here. 

Mass loss in collisions of main-sequence stars has been already explored to a considerable extent, for example, \cite{dale} found that in encounters that involve a massive main-sequence star and a low mass one, the mass loss is between 2-4\%. \cite{gaburov} studied mixing in massive stellar mergers, and the highest mass loss they obtained was 8.9\%. \cite{gleb} modelled central collisions between low mass stars, obtaining mass losses of a few percent. 

We present here an investigation which explores the effect of mass loss during collisions, on the formation of massive black hole seeds from a dense stellar cluster. The numerical
methods used are described in Section 2, including our setup and the mass loss parametrizations, and the main results are presented in Section 3. We summarize our work and address the main conclusions
in Section 4.

\begin{figure*}
\centering
\includegraphics[trim={10cm 1cm 10cm 0},clip, width=1.\textwidth]{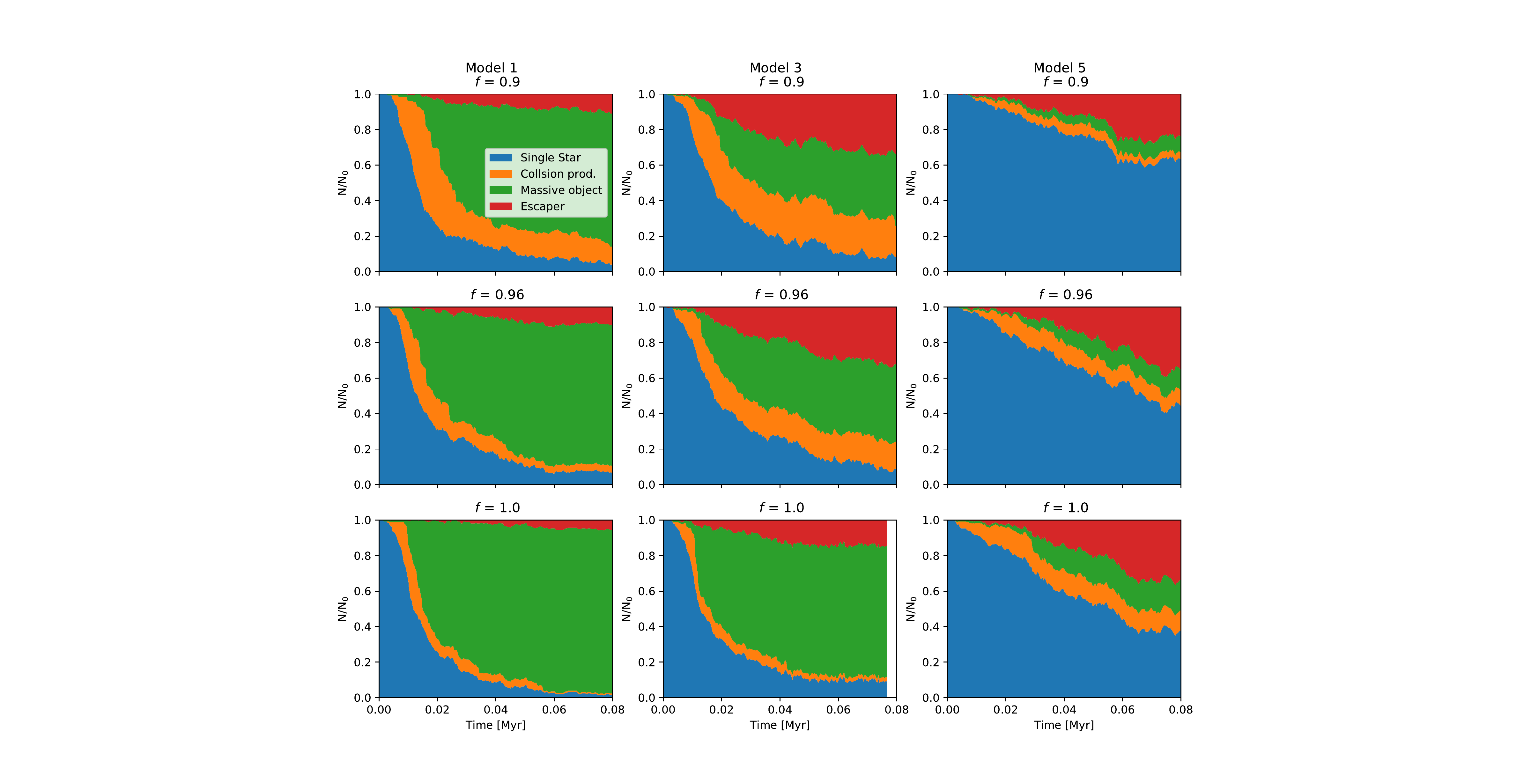}
\caption{Fraction of stars over time corresponding to four categories: escapers (red);  stars that have collided with and are part of the most
massive star in the system (green, ‘Massive object’); stars that are part of other collision products (orange, ‘Collision prod.’); and single stars (blue), for different models, and different values for the fraction of mass conserved after a collision $f$. The general trend for the models is that the fraction of stars being part of the most massive object decreases when $f$ decreases.}
\label{fig:ms}
\end{figure*}

\begin{figure*}
\centering
\includegraphics[trim={9cm 1cm 10.2cm 0},clip,width=.91\textwidth]{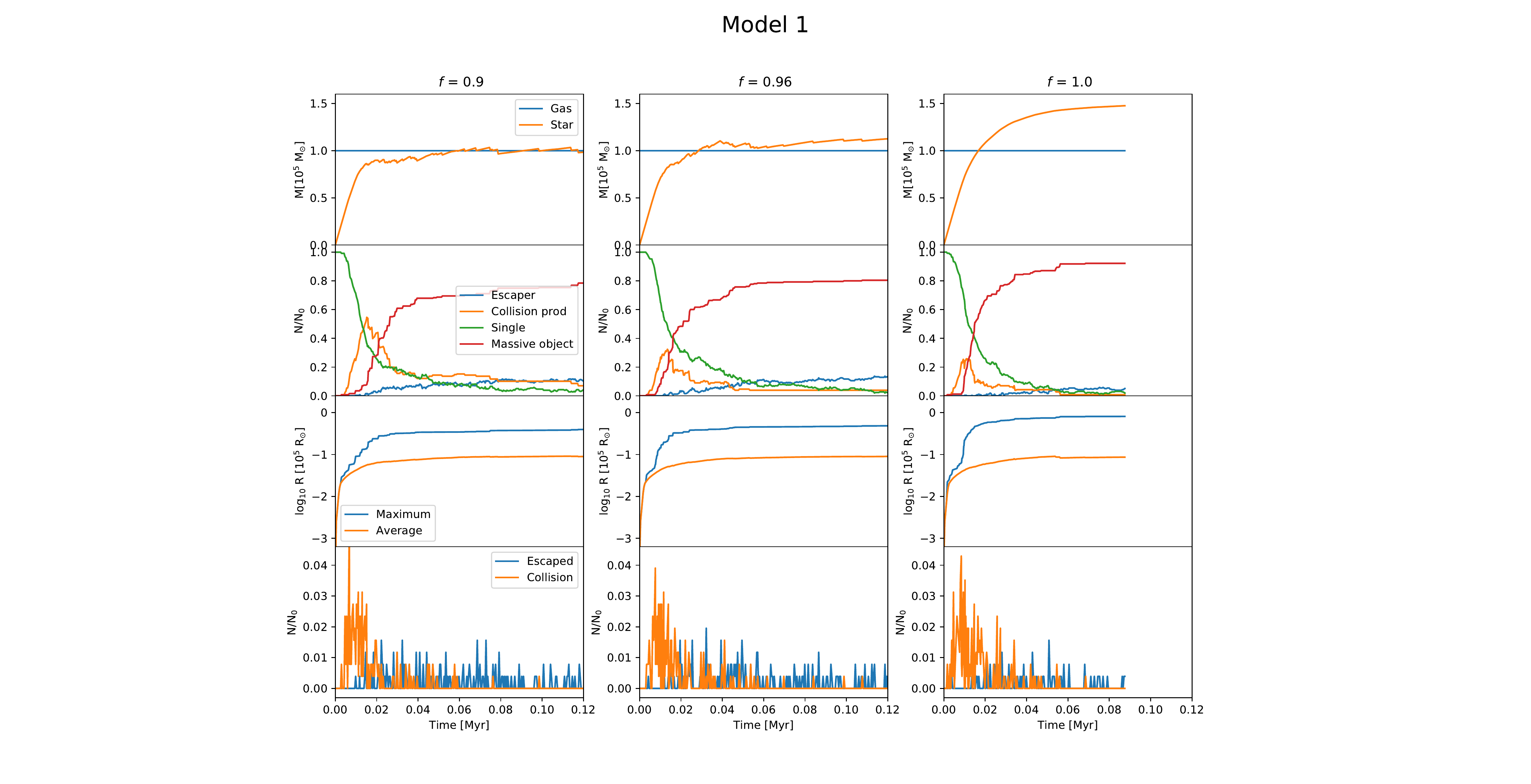}
\caption{Correlation of  the time evolution of the collision and escape rate (bottom panel) with: total star and gas mass (top panel), fraction of stars belonging to
the same four categories as in Fig. \ref{fig:ms} (second panel), and maximum stellar radius and average stellar radius of the remaining stars (third panel), for three different and representative values of the retained mass fraction $f$.}
\label{fig:m1}
\end{figure*}

\begin{figure*}
\centering
\includegraphics[trim={9cm 1cm 10.5cm 0},clip, width=.91\textwidth]{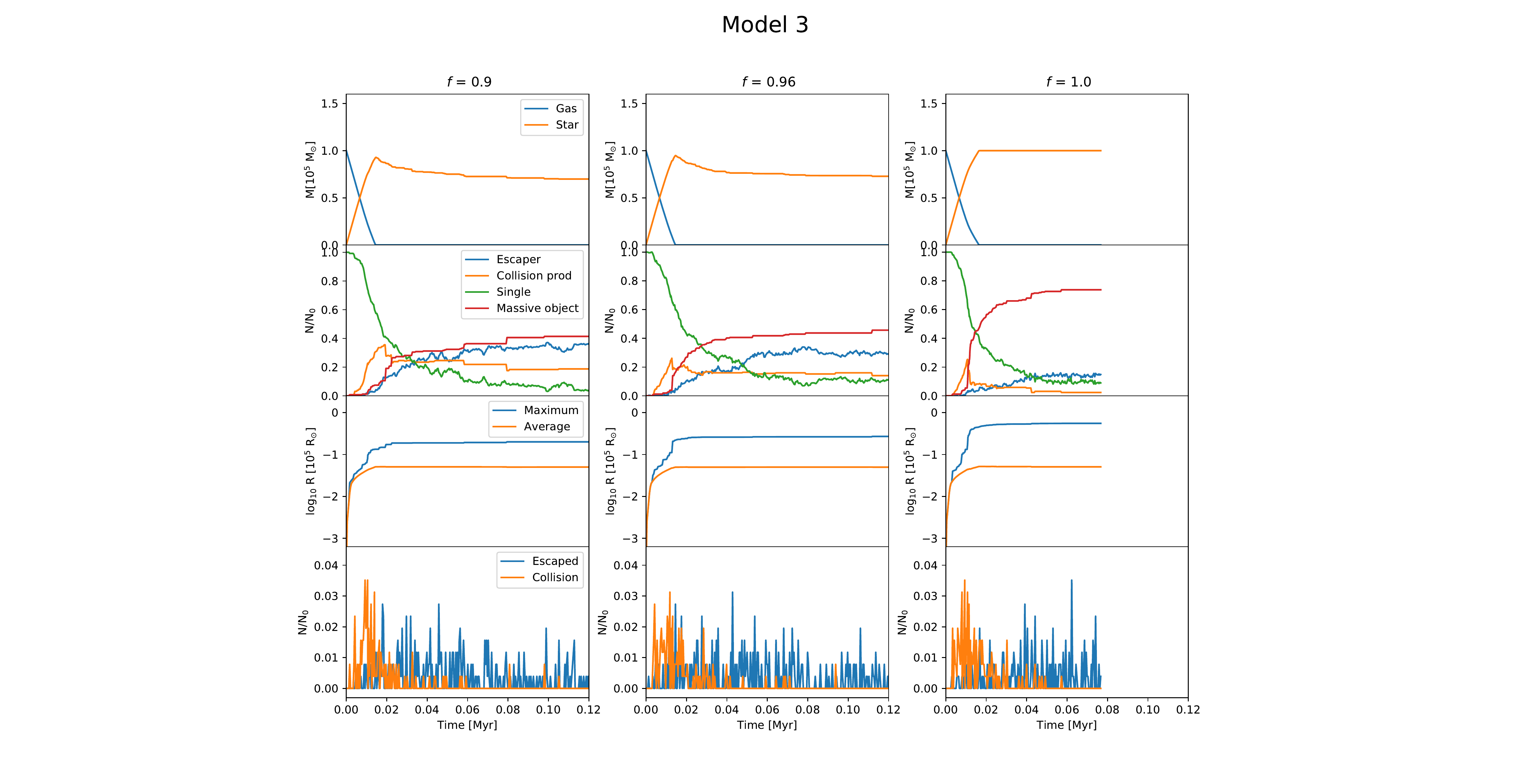}
\caption{Same as Fig. \ref{fig:m1} for model 3.}
\label{fig:m3}
\end{figure*}

\begin{figure*}
\centering
 \includegraphics[trim={9cm 1cm 10.5cm 0},clip, width=.91\textwidth]{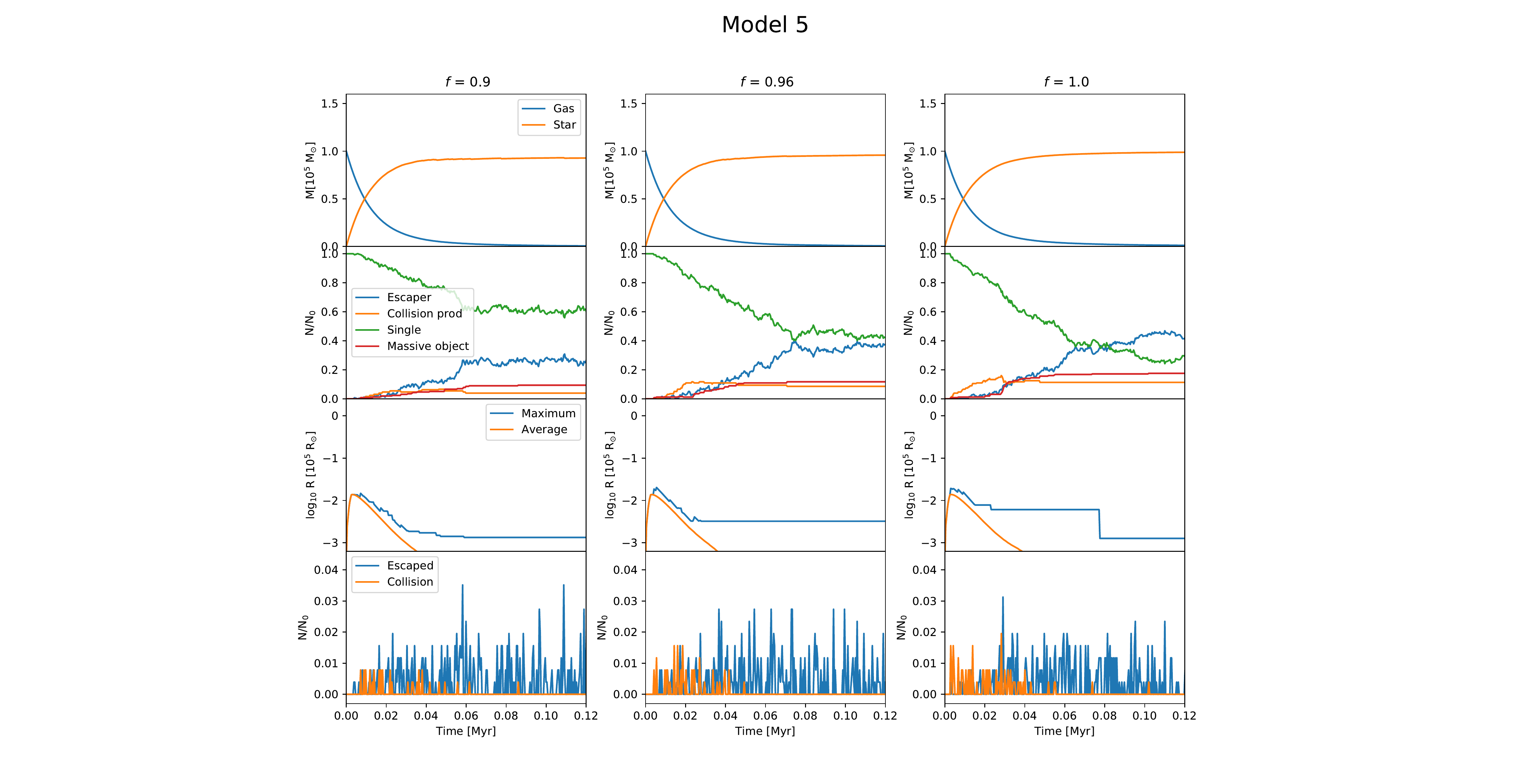}
\caption{Same as Fig. \ref{fig:m1} for model 5.}
\label{fig:m5}
\end{figure*}

\section{Numerical Methods}
\label{sec:2}
In order to model our astrophysical system, which  consists of Pop. III protostars embedded in their natal gas cloud, we consider a variety of physical processes that play a role. Besides the gravitational $N$-body dynamics of the cluster, we model the gravitational coupling between the gas and the stars, include gas accretion which leads to stars growing in mass and size, and also handle collisions between stellar components of the system.

We use the Astrophysical MUlti-purpose Software Environment
\citep[AMUSE,][]{amusebook,port09,port13,pelup}, designed to perform multi-physics simulations, which we require in this investigation. AMUSE is particularly helpful, because it allows us to introduce  a mass-radius parametrization for accreting Pop. III
protostars, and to couple it to existing $N$-body codes.

We
assume that the protostars and the gas are distributed equally, both following a Plummer distribution, where gravitational dynamics of the gas cloud are included by using  an analytical background potential. The protostars will gain mass
by accreting from the gas, and their radii are completely determined by
the mass  and accretion rate, at every time step in
our simulation. 
The star-star gravitational interactions are modelled using the
$N$-body code ph4. The Bridge scheme \citep{fuji07} is used to couple the stars to the gas potential, so that the stars experience the gravitational force from each other as well as from
the gas.

    In order to treat collisions between protostars we use the 'sticky-sphere' approximation: whenever the
distance between two protostars is less than the sum of their
radii, we replace the two protostars by a single object at the center of mass, with a new mass, and new radius determined by the mass-radius parametrization.

 To determine the value of the mass of this new single object, we have to account for the mass loss in the collision. This mass loss fraction should take into account the most important parameters of a collision process. We describe the final mass of the collision product as follows:
 \begin{equation}
     M_t= (M_1+M_2)\times f(M_1,M_2,R_1,R_2),
     \label{eq:mass}
 \end{equation}
     where $f$ is a function that regulates the mass loss effect, which we assume here to depend mostly on the stellar mass and the stellar structure, based on current results in the literature. It is conceivable that also the collision velocity may play a relevant role as well as the difference in the dynamics between two-body and three-body mergers, but these need to be explored in more detail via hydrodynamical simulations. It has however been shown in simulations by \cite{gaburov2010} that the hydrodynamics of three-body collisions (a binary and a third star) are well-described with the sticky-sphere approximation plus additional mass loss in at least 75\% of all cases. We assume here that the mass being ejected during the merger will be ejected at high velocities, and consequently escape from the gravitational potential. It is therefore not being returned to the gas reservoir.
    
   The literature on stellar mergers tells us that in a collision the mass loss depends on the stellar evolutionary stage (weakly bound envelopes in older stars), mass and collision parameters, and the final product retains between 90\% and 100\% of the total mass \citep{dale,gaburov,gleb}. So as a first approximation, we assume constant values for the mass loss ($f$ is the fraction kept in collisions), and see how a constant mass loss fraction in every collision affects the final mass of the most massive object formed in the cluster. If the final object is massive enough, i.e. reaching at least 1000 $M_\odot$, we expect that it will evolve into a massive black hole at the end of its lifetime. So far, simulations of stellar mergers have focused on collisions between main-sequence stars and evolved stellar objects, while collisions between protostars have been explored to a lesser degree in the context of the formation of massive stars \citep[see e.g.][]{Baumgardt_2011}. To our knowledge, mass loss in protostellar collisions has not been studied. The latter would be important to better understand the implications of such mergers in black hole formation scenarios as considered here. In the absence of further information, we here adopt the parametrizations derived by \cite{lomb02} and \cite{gleb} for the mass loss, combining them with approximate protostellar models.

The parameters that specify the initial
conditions in the simulations are the total gas mass, $M_g$; the cut-off radius of the gas cloud, $R_g$; the number of protostars, $N$; and the average accretion rate, $ \dot{m}$.
We begin our simulations considering the 6 different accretion scenarios defined in \cite{tjarda}, which are based on:
\begin{itemize}
 \setlength\itemsep{0cm}
    \item Finite or infinite
gas reservoir, where the second resembles a system that is constantly being fed fresh gas, contrary to
the finite gas reservoir models, where we remove the accreted gas from the reservoir, thereby depleting the reservoir. Once it is fully depleted, the accretion rates are set to zero.
\item Position dependent accretion rates, where the accretion rate is set proportional to the local gas density. In this way the protostars in
the core accrete at a higher rate than protostars in the halo.  
\item Time dependent accretion rates, where we further assume the accretion rate to be proportional to the remaining gas reservoir. We start with the same initial rates, which are decreased as gas is depleted from the reservoir.
\end{itemize}
 The 6 different models are summarized in Table \ref{tab:tjar}. 
\begin{table}
\caption{Different accretion models studied.}
  \label{tab:tjar}
  \begin{center}
\begin{tabular}{ cccc }
\hline
 Model & Gas  & Position dependent & Time dependent  \\
   &  reservoir &  accretion model & accretion model \\
 \hline
 1 & Infinite & No & No \\  
 2 & Infinite & Yes & No \\  
 3 & Finite & No & No \\  
 4 & Finite & Yes & No \\  
 5 & Finite & No & Yes \\  
 6 & Finite & Yes & Yes \\  
 \hline
\end{tabular}
\end{center}

\end{table}
We also consider the standard set of parameters used in \cite{tjarda}, which are: $M_g=10^5 M_{\odot} $, $R_g=0.1$ pc, $N = 256$, $ \dot{m} = 0.03 M_{\odot}$yr$^{-1}$, and a mass-radius parametrization based on \cite{hoso}. The initial mass of all the protostars
is set to $m_0 = 0.1 M_{\odot}$ (We note that the mass of the protostar subsequently starts to evolve quickly due to the accretion recipes as well as mergers). The selection of this particular set of values indicates that we are interested in very massive Pop. III protostar clusters that could lead to the formation very massive objects. We also varied the factor $f$ between 0.85 - 1.00 in steps of 0.01. In order to have good statistics we performed 5 simulations for each value of $f$ per model through models 1, 2, 3, 4 and 6, and 3 simulations for model 5.

As we work with a constant mass loss, we have to consider that in the case of a collision of a very massive object with a light one, the sum of their masses multiplied by a constant factor smaller than 1 could give a final mass smaller than the mass of the most massive collision component.  We decided to perform two types of simulations, one that allows for a decrease in the total mass after a collision and also one that just allows for the mass to increase. In the second case, if the mass of the collision product is less than the previous mass, we decided to set the collision product mass equal to the mass of the most massive collision component.  

We determine if a simulation is finished by keeping track of the average collision rate $$\nu_{{av}}(t)=\frac{N_{{col}}(t)}{{t_{{last \,collision}}}}$$
and an upper limit of the current collision rate,
$$\nu(t)=\frac{1}{{t-t_{{last \,collision}}}}\hspace{.05cm}.$$
If the ratio of $\nu/\nu_{{av}} <$ 0.015 the simulation stops. This will ensure that most of the collisions have occurred.

After assuming a constant $f$, our aim was to find a more realistic prescription for the mass loss fraction. \cite{lomb02} presented simulations of stellar collisions using a SPH-code, providing the following prescriptions to fit the mass ejected by the collision:
   
 \begin{equation}
     \phi = C_1\frac{q}{(1+q)^2}\frac{R_{1,0.86}+R_{2,0.86}}{R_{1,0.5}+R_{2,0.5}} \hspace{.1cm}, \label{eq:gleb1}
 \end{equation}
 where $\phi$ is the fraction of mass ejected, $C_1=0.1574$, $q$ is the mass ratio $M_2/M_1$, and $R_{n,0.5}$, $R_{n,0.86}$ are the radii containing 50 and 86 percent of mass of the parent star $n$ (1 or 2). \cite{gleb} found that when the stellar structures are more equal, the mass loss could also be modelled using the simpler prescription
 \begin{equation}
     \phi = C_2 \frac{q}{(1+q)^2}
     \label{eq:gleb2}
 \end{equation}
with $C_2=0.3$. 

 From \cite{dom} we get a mass-radius relationship for accreting primordial protostars:
   \begin{equation}
       \frac{1000R_{\odot}}{R}=\frac{1000}{260\,(M/M_{\odot})^{1/2}}+\frac{1.04\,\mathrm{yr}^{-1}(t-t_{ini}(M))}{M/M_{\odot}}.
       \label{eq:dom}
   \end{equation}
 Each term on the right side of the equation depends on a characteristic timescale of the protostar. If we want to determine the radius containing the decimal percentage $i$ of the mass of the parent star $n$, we have two cases depending which timescale dominates:
 
 If the timescale for accretion is much larger than the timescale for protostellar contraction, i.e. $t_{acc}\gg t_{KH}$ the first term dominates and we have
     \begin{equation}
         R_{n,i}=260\left(\frac{i\cdot M_n}{M_{\odot}}\right)^{1/2} R_{\odot}.
         \label{eq:tim1}
     \end{equation}
      On the other hand, if $t_{KH}$ dominates
     \begin{equation}
         R_{n,i}=\frac{i\cdot M_n}{1.04 \times 10^{-3} M_{\odot}}\frac{1}{t_{KH}/ \mathrm{yr}  }R_{\odot}.
         \label{eq:tim2}
     \end{equation}
 The relevant timescales are given as: 
 $$t_{acc} = \frac{M}{\dot{M}}\hspace{.2cm},  \hspace{1cm}    t_{KH}=\frac{GM^2}{RL}\hspace{.2cm},  $$
 where we assumed $L$ to be the Eddington luminosity, $L_{Edd} = 3.8\times10^4L_{\odot}(M/M_{\odot})$. We introduced this prescription for the mass loss in the simulations, and performed 5 simulations per accretion model.



\section{Results}
\begin{figure}
\centering
\includegraphics[width=1.\columnwidth]{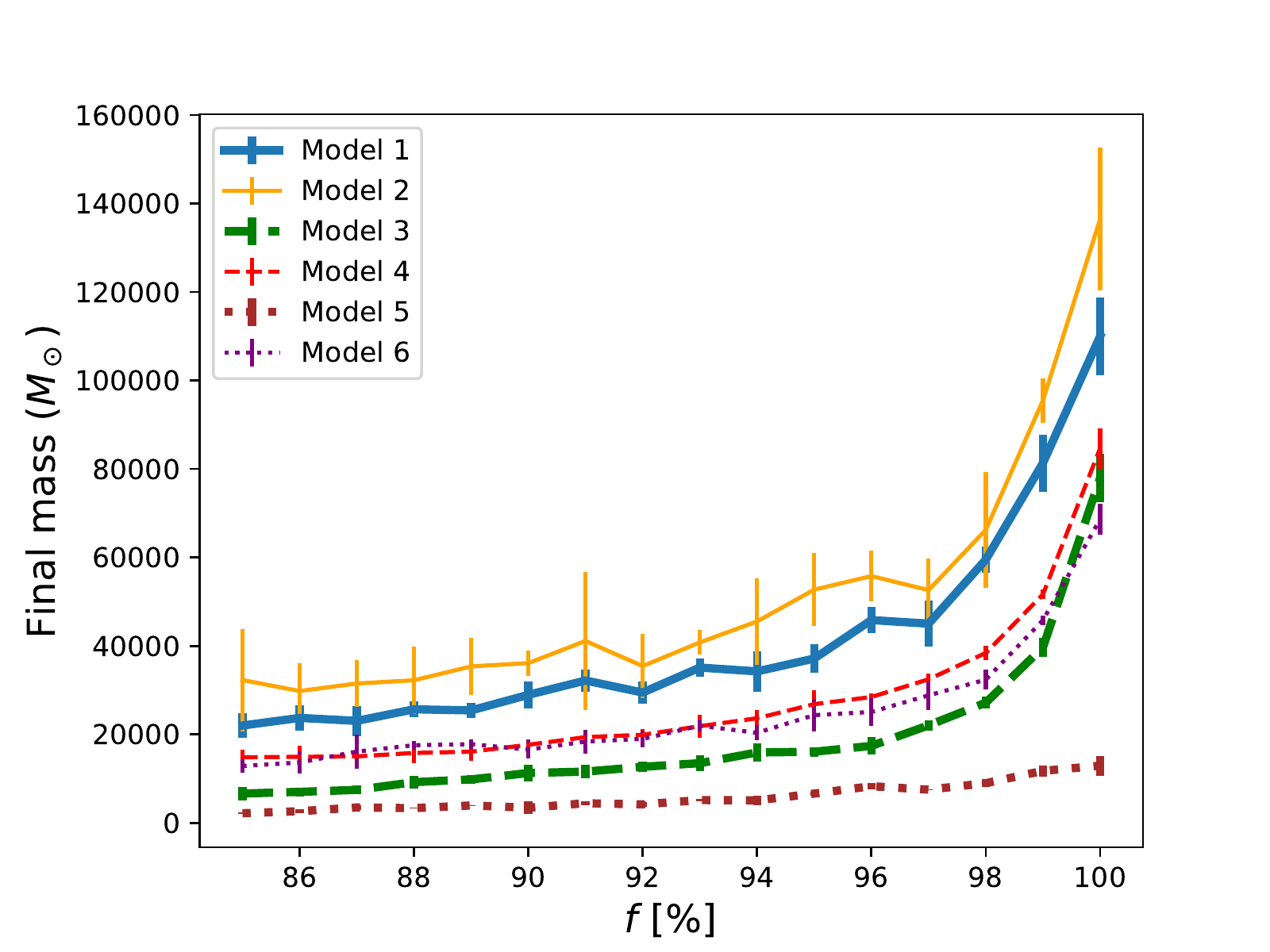}
\caption{Final mass of the most massive object at the end of each simulation, as a function of the retained mass fraction $f$, for each model described. The bar at each point simulated represents the $\sigma$ error considering 5 simulations for each $f$ per model.}
\label{fig:c}
\end{figure}

\begin{figure*}
  \centering
  \begin{minipage}[!h]{0.48\textwidth}
    \includegraphics[width=1.\textwidth]{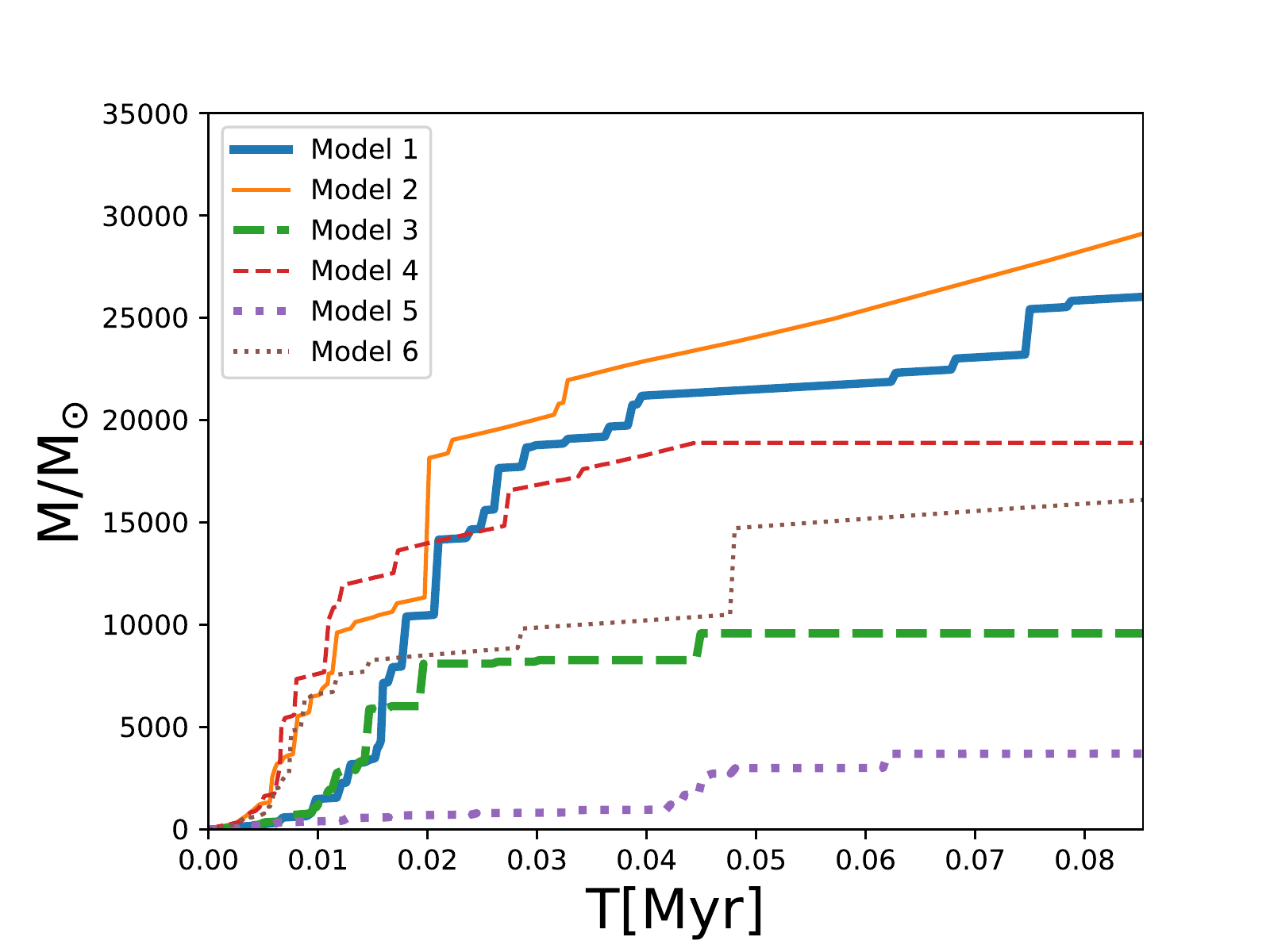}
    \caption{Time evolution of the mass of the central object, for six different accretion models and a 10\% mass loss per collision. All models, except for number 5, efficiently convert gas mass into one massive object.}
    \label{fig:asd1}
  \end{minipage}
  \hfill
  \begin{minipage}[!h]{0.48\textwidth}
    \includegraphics[width=1.\textwidth]{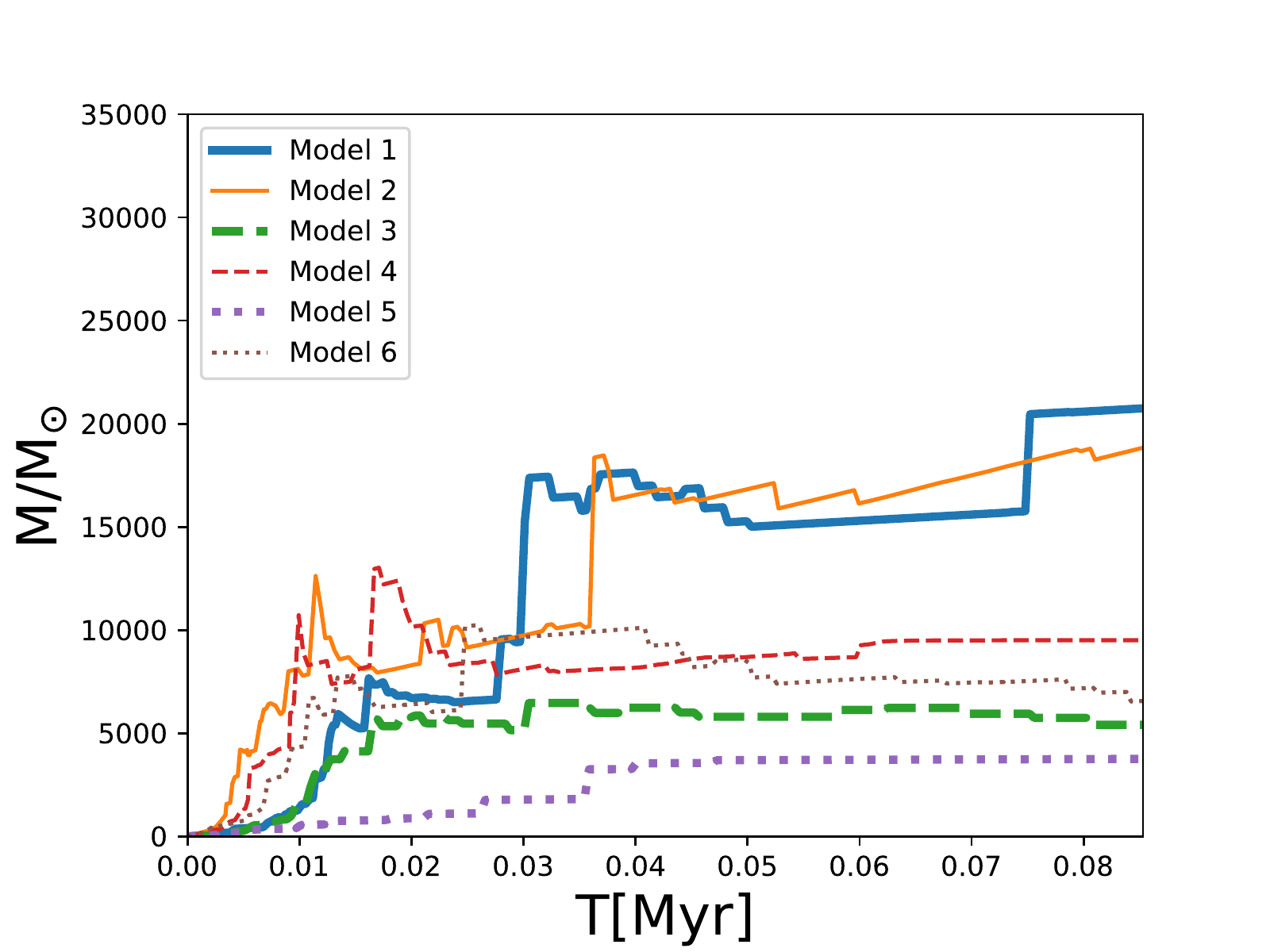}
    \caption{Same as Fig. 8, but in a scenario that allows for a decrease in the mass after a collision. Models 1 and 2 efficiently convert gas into one massive object, the efficiency of the other models is lower than in the other scenario. }\label{fig:asd2}
  \end{minipage}

\end{figure*}

\begin{figure*}
\includegraphics[trim={11cm 1cm 11cm 0},clip,width=\textwidth]{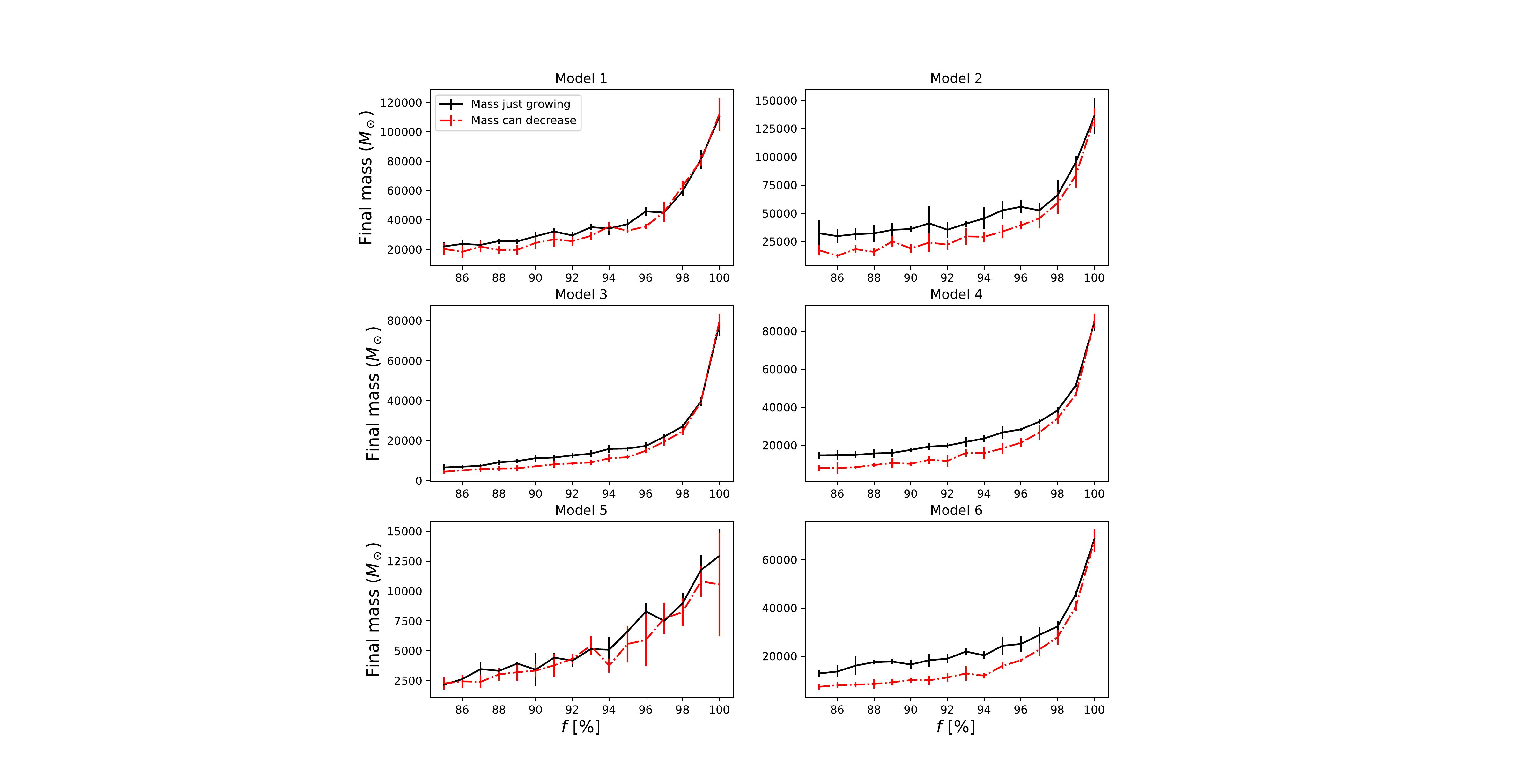}
\centering
\caption{Comparison of models where the mass of the most massive object can only increase (black line) after a collision, and models that allow a decreasing mass (red). Each point in each plot has a bar representing the $\sigma$ error considering 5 simulations for each retained mass fraction $f$ per model.}
\label{fig:a}
\end{figure*}

The main goal of our simulations is to better understand how the mass of the central object in the cluster changes when mass loss is considered. In the first simulations, we considered constant values for the factor $f$ that regulates mass loss during collisions, and studied how the cluster evolves.

The simulations start with a cluster of single protostars. As the system evolves, different collision products will form. Also, dynamical encounters among protostars can eject them from the cluster. Thus we define four categories to which a protostar in the simulation can be part of: single protostar: which are part of the cluster but are not part of a collision product, the most massive collision product, a less massive collision product, and an escaper: which are stars that are far away from the cluster with positive energy.
  Figure \ref{fig:ms}  presents the time evolution of the fraction of
protostars belonging to each of the defined categories, for the position independent models (1, 3 and 5), which represent the different kind of behaviors for every model. The majority of protostars end up in the final most massive object for models 1 and 3, while most of the stars do not undergo merger events in model 5. In the first models, as a result of gas accretion, the protostars are growing in radius, because protostellar
mass is also increasing, thus getting a larger cross-section, resulting in a high collision probability in this environment. As we decrease the value of $f$, the fraction of objects that become part of the most massive object decreases, but the number of protostars that escape, or take part in a less massive collision product increases, as a result of the mass. In model 5, because of the uniform accretion, the central object in this model is less massive, and due
to the time-dependent accretion, the protostellar radii shrink again as the cluster runs out of gas. We observe that most of the stars remain single, and we see a higher fraction of escapers than in the other models; as we decrease $f$, the fraction of single stars becomes higher while the other fractions decrease. 
We can also say that inflated radii turn ejection events into collisions.

\begin{figure*}
\centering
\includegraphics[trim={11cm 1cm 11cm 0},clip,width=\textwidth]{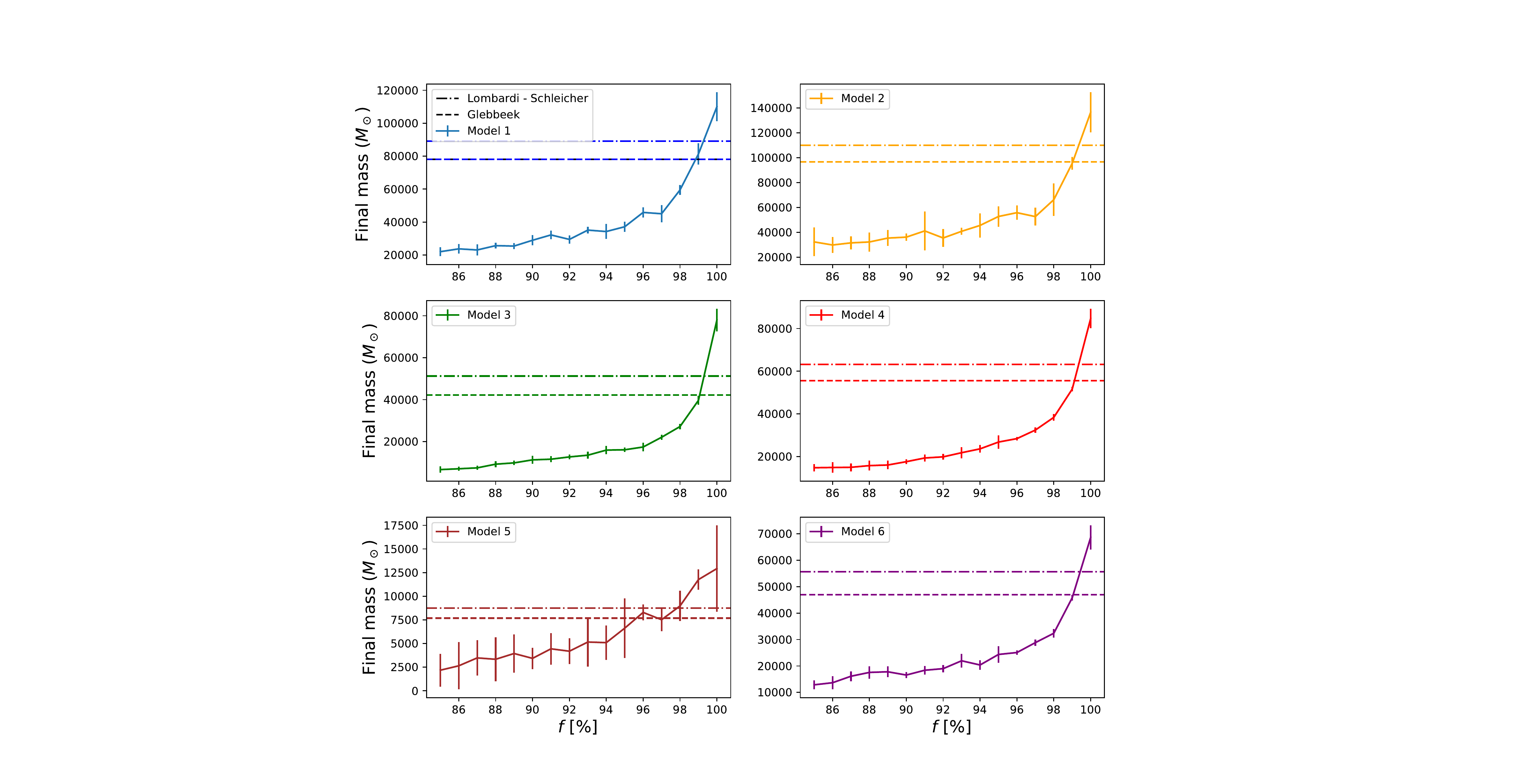}
\caption{Results of simulations using the mass loss prescriptions by Lombardi and Glebbeek, compared with the mass loss just considering a constant retained mass fraction f $f$. The dashed line in each plot represents the simpler prescription given by equation (2), meanwhile the dash-dotted line represents the results of the more complex parametrization using the Schleicher et al. (2013) mass-radius relationship and the equation (1). Each line represents the mean of final mass for 5 simulations. }
\label{fig:b}
\end{figure*}

 Figures \ref{fig:m1}, \ref{fig:m3}, and \ref{fig:m5}, present the collision
rate (bottom panels) and correlate it with the total star and gas mass
(top panels), the fraction of stars belonging to the four categories
defined earlier (second panels), and the radius of the most massive
protostar and the average radius of all the remaining protostars (third
panels), for the three position independent models 1, 3, and 5, using $f$ = 0.9, 0.96 and 1. From \cite{tjarda} the initial crossing time for a system with our parameter values is 853 yr, and in every model the collision rate starts to increase rapidly on a timescale of the order of 10$^4$ years after the simulation has begun, which is equivalent to nearly a dozen crossing times, as well as a short time span compared to the total duration of the simulation. This timescale corresponds to the time it takes for the protostars to gain mass and obtain a larger radius, and also for the total stellar mass to be comparable to the gas mass. After reaching its peak, the collision rate decreases steadily and a small amount of objects starts to escape. The protostars grow rapidly and make the system stellar-mass dominated, producing dynamical encounters and collisions. For model 1 we find that the stars have a tendency to form a massive central object, with a low rate of escapers through the simulation. Model 3 presents an equally efficient way to form a massive object, but the rate of escapers is higher, especially at later times. Model 5 forms an object of moderate mass, and presents a steady escape rate.

The comparison between the scenarios with different values of $f$ is clear. A lower value of $f$ corresponds to a lower value in the total stellar mass of the system, and a lower fraction of the number of stars that form the most massive object in the cluster, also increasing the number of objects that escape. The average radius of the stars does not seem to be affected, but it is clear that the radius of the most massive object decreases with $f$. The largest difference is seen between $f$ values of 0.96 and 1. For 0.9 and 0.96, while there are differences, they are more subtle and not as important, which suggests that in this range of higher mass loss the simulations evolve in a similar way, so changing the value of $f$ is not as fundamental in this regime.

Now, to understand the effect on the final mass of the central object, we show the final mass as a function of $f$ in Fig. \ref{fig:c}. We find that for models 1 and 2 the behavior of the curve changes abruptly when $f$ reaches a value of $\sim$0.96, for models 3, 4 and 6 there is also a sudden change of behavior around the same value. However, it is not as steep as the first one, and for model 5 there is no abrupt global change in the behavior of the plot, this can be seen in the Fig \ref{fig:c}. With these observations we can say that there is a steep dependence on $f$ in the interval $f=[0.96-1.00]$, where the mass of the final object changes extremely with the value of $f$, so that even a small mass loss fraction is still relevant, obtaining values of almost one third of the mass obtained in simulations where mass loss is not considered. On the other hand, in the range $f=[0.85,0.96]$, changes in the value of $f$ do not lead to a big change in the final mass obtained. This suggests that, when considering a constant mass loss per collision, there is a breaking point where the mass loss fraction is a deciding factor in the resulting final mass of the most massive object. 

In order to explain the change of behavior when we approach values of $f$ of $\sim 0.96$, we note that the typical masses of the protostars, around the final stages when they collide, are between $500-1000M_{\odot}$, while the central object has a mass of the order of several $10^5M_{\odot}$, therefore the mass loss is comparable to the mass gain during the collision, reducing the net effect. Considering the accretion rate, mass increases by about $300M_{\odot}$ in 0.01 Myr, so for this range of values of $f$  the mass loss during collisions with the massive object becomes comparable to the mass that has previously been accreted. We can say that for high values of $f$, the growth of the mass is dominated by collisions, and for lower values, it is dominated by accretion.

Also, we contrast scenarios where the mass of the most massive object can only increase after a collision, and scenarios that allow for a decrease. First, we compare the efficiency of the models to produce a massive object on a given timescale for a constant mass loss of 10\%, as this represents a lower limit in the literature, for both scenarios. Figures \ref{fig:asd1} and \ref{fig:asd2}, show the growth of the most massive object for these scenarios, demonstrating that when allowing for the mass to decrease after a collision, the efficiency to create a massive object decreases. We show the dependence on $f$ for the final mass for both scenarios, for each model, in Fig. \ref{fig:a}. Due to the nature of our mass loss equation \ref{eq:mass}, when $f$ approaches 1 the mass we are going to obtain is the same as if we do not consider mass loss at all in the simulations, regardless of the kind of scenario we are in, so it is to be expected that both types of simulations behave the same at the higher end of the plot. What we see is that both types of simulations behave almost exactly the same for most of the models, for values of $f$ in the interval [0.96, 1]; both kind of simulations give similar final masses. This coincides with the range in which a small change in $f$ becomes relevant for the final mass. In particular we note that in models 1 and 3 there is little difference between the scenarios across all the values of $f$ tested, while in models 2, 4 and 6 the difference between these scenarios is more notorious for values of $f$ lower than 0.96, and  model 3 exhibits a more erratic behavior at the higher end of the plot, but the masses obtained for both models are in the same range. This tells us that in the range where changes in $f$ are crucial, the results do not sensitively depend on the type of model, but for lower values of $f$, there are distinguishable differences, particularly for position dependent accretion models, as that in this regime the infrequent collisions may even remove some of the mass that has been accreted before. 

Finally, we compare the results of the simulations with the parametrizations of \cite{lomb02} and \cite{gleb}. These parametrizations provide a single value for the mass obtained for each model, and we compare this value with the results of the simulations employing constant mass loss fractions, so we could see in which range we can consider the prescriptions to be equivalent to a constant mass loss scenario. Figure \ref{fig:b} shows the results we get for both analytical models, compared with the result we previously had in the constant mass loss scenario, for each accretion model. In general, the trend is that the prescriptions can be compared to a constant mass loss configuration where $f\sim 0.99$, nonetheless the exact value differs with the accretion model we consider. We also note that the parametrization using the \cite{dom} protostellar models gives slightly higher values for the final mass than the simpler prescription by \cite{gleb}, and the dispersion between them depends also on the accretion model we consider.  The final mass values obtained from these simulations, including the calculated value of $f$ that would give that same mass in a constant mass loss scenario, and the corresponding percentage of the mass it represents when compared to models without mass loss are shown in Table \ref{tab:gleb}.
\vspace{-0.2cm}
\begin{table}
    \caption{Overview of the results using the Lombardi and Glebbeek parametrizations, combined with the protostellar structure model by Schleicher et al. (2013) , including final mass, calculated $f$ value, and percentage of the mass compared to a scenario without mass loss. }
    \label{tab:gleb}
\begin{tabular}{ c c c c c c c}

\hline
 &\multicolumn{3}{|c|}{Glebbeek} & \multicolumn{3}{|c|}{Lombardi - Schleicher}\\
\hline
 Model  & Mass ($M_{\odot}$) & $f$ & \% & Mass ($M_{\odot}$) & $f$ & \% \\
 \hline
 1 & 78003 & 98.84 & 74 & 89151 & 99.33 & 85\\  
 2 & 96714 & 99.03 & 70 & 109889 & 99.35 & 80\\  
 3 & 42214 & 99.06 & 54 & 51240 & 99.30 & 65\\  
 4 & 55610 & 99.12 & 65 & 63185 & 99.34 & 74\\  
 5 & 7683 & 97.11 & 59 & 8737 & 97.83 & 67\\  
 6 & 46980 & 99.05 & 68& 55595 & 99.42 & 81\\  
 \hline
\end{tabular}
\end{table}
\vspace{-0.2cm}
\section{Summary and Conclusions}
We have presented a study of how the mass loss during collisions can affect the final mass during the formation of SMBH seeds, formed through collisions and accretion, in a dense, primordial Pop. III protostellar cluster. We take into account standard cluster parameters as introduced by \cite{tjarda}, which ensure the production of massive black hole seeds. We investigated the effect of a constant fraction of mass loss per collision, for later comparison with more complex prescriptions. For that, we performed a
series of simulations using the AMUSE framework, which included $N$-body dynamics, an
analytical gas potential, different accretion models, mass-radius
parametrizations, and stellar collisions.

First, we present the time evolution of the stellar components of the cluster for representative accretion models in Figure \ref{fig:ms}, which shows that for higher mass loss, the fraction of stars which become part of the most massive object decreases, while the single stars and escaper fraction gets larger, due to the fact that the cluster gravitational potential becomes smaller when mass loss increases.  

In Figure \ref{fig:c} we present the results of our first simulations, showing the dependence of the final mass of the most massive object with the factor $f$ that describes mass loss. We can clearly see two distinct behaviors, for higher values of $f$ the mass depends steeply on its value,  while for lower values the final mass does not change much with  $f$, this can be explained because at the higher end of the plot the growth of the mass is dominated by collisions, and accretion dominates for the lower end. Figure \ref{fig:a} compares scenarios where the mass of the most massive object can only increase after a collision and scenarios that allow for a decrease, where we see that the differences between scenarios are more significant for lower values of $f$, while for higher values there is not much difference. This shows that in the range where changes in $f$ are important, whether or not we allow for the mass of an object to decrease after a collision is not really relevant. In this context, a constant mass loss of 5\%  represents a final mass between 60-80\% lower (50\% for model 5) than the mass we would get if we did not consider mass loss in our models. 

Also, we compared the results of the constant mass loss  simulations with more complex mass loss parametrizations by \cite{lomb02} and \cite{gleb}, using the \cite{dom} mass-radius relationships for primordial protostars. The values obtained tell us that these analytical models are equivalent to scenarios of constant mass loss of $\sim$1\% per collision. However, this low mass loss percentage can have a great impact on the final mass of the object, that could lose between 15 to 40\% of its mass depending on the accretion model we study. 
Considering this, even in the most extreme mass loss scenarios, we get final masses of the order of $10^4M_{\odot}$, which can be considered massive enough to be SMBH seeds, confirming  that the model is still a viable candidate for explaining the SMBH we see nowadays.

The mass loss in collisions of primordial protostars has not been studied, and given the importance of mass loss in the context of the formation of SMBH seeds, it is important to explore this in more detail in the future via hydrodynamical simulations, considering also three-body mergers and different relative velocities, with the aim to derive approximate relations that can subsequently be employed in $N$-body models. More precise models will help us better understand how we can really form the massive black holes we see today. 
\section*{Acknowledgements}

PJAS and DRGS thank for funding via Fondecyt regular 1161247 and Conicyt PIA ACT172033.
TB acknowledges support from ENGAGE SKA RI, grant POCI-01-0145-FEDER-022217, funded by COMPETE 2020 and FCT, Portugal and from grant UID/EEA/50008/2019 funded by FCT. RSK receives financial support from the Deutsche Forschungsgemeinschaft (DFG, German Research Foundation) -- Project-ID 138713538 -- SFB 881 (``The Milky Way System'', subprojects A1, B1, B2), and he acknowledges support fom the DFG via the Heidelberg Cluster of Excellence {\em STRUCTURES} in the framework of Germany's Excellence Strategy (grant EXC-2181/1 - 390900948). MF was funded via Fondecyt regular 1180291.





\bibliographystyle{mnras}
\bibliography{bib}



\appendix

\section{}
\begin{table*}
    \caption{Average final masses of the most massive object, obtained in simulations for constant $f$, and the percentage they represent of the most massive result }
    \label{tab:asd}
\begin{tabular}{ c c c c c c c c c c c c c }

\hline
  &\multicolumn{2}{|c|}{Model 1} &\multicolumn{2}{|c|}{Model 2} &\multicolumn{2}{|c|}{Model 3} &\multicolumn{2}{|c|}{Model 4} &\multicolumn{2}{|c|}{Model 5} &\multicolumn{2}{|c|}{Model 6}\\
\hline
$100f$ & Mass ($M_{\odot}$) & \% & Mass ($M_{\odot}$) & \% & Mass ($M_{\odot}$) & \% & Mass ($M_{\odot}$) & \% & Mass ($M_{\odot}$) & \% & Mass ($M_{\odot}$) & \%  \\

 \hline
85 & 22009 & 21 & 32290 & 23 & 6627 & 8 & 14840 & 17 & 2178 & 16 & 12862 & 18  \\ 
86 & 23731 & 22 & 29803 & 21 & 7021 & 9 & 14927 & 17 & 2642 & 20 & 13652 & 19  \\ 
87 & 23059 & 22 & 31503 & 23 & 7456 & 9 & 15027 & 17 & 3479 & 26 & 16112 & 23  \\ 
88 & 25666 & 24 & 32230 & 23 & 9209 & 11 & 15810 & 18 & 3332 & 25 & 17559 & 25  \\ 
89 & 25428 & 24 & 35371 & 25 & 9826 & 12 & 16099 & 19 & 3943 & 30 & 17791 & 25  \\ 
90 & 28942 & 27 & 36110 & 26 & 11274 & 14 & 17651 & 20 & 3429 & 26 & 16561 & 24  \\ 
91 & 32182 & 30 & 41123 & 30 & 11586 & 14 & 19384 & 22 & 4439 & 34 & 18374 & 26  \\ 
92 & 29457 & 28 & 35436 & 25 & 12674 & 16 & 19912 & 23 & 4186 & 32 & 18982 & 27  \\ 
93 & 35107 & 33 & 40799 & 29 & 13496 & 17 & 21854 & 25 & 5157 & 39 & 21958 & 32  \\ 
94 & 34242 & 32 & 45533 & 33 & 15921 & 20 & 23655 & 27 & 5092 & 39 & 20362 & 29  \\ 
95 & 37147 & 35 & 52715 & 38 & 16082 & 20 & 26838 & 31 & 6621 & 51 & 24351 & 35  \\ 
96 & 45838 & 43 & 55783 & 40 & 17400 & 22 & 28448 & 33 & 8287 & 64 & 25074 & 36  \\ 
97 & 45031 & 43 & 52618 & 38 & 22053 & 28 & 32457 & 38 & 7515 & 58 & 28820 & 42  \\ 
98 & 59486 & 56 & 66223 & 48 & 27167 & 34 & 38402 & 45 & 8971 & 69 & 32367 & 47  \\ 
99 & 81334 & 77 & 95491 & 69 & 39667 & 50 & 51641 & 60 & 11760 & 90 & 45806 & 66  \\ 
100 & 104674 & 100 & 136457 & 100 & 77962 & 100 & 84671 & 100 & 12925 & 100 & 68615 & 100  \\ 

 \hline
\end{tabular}
\end{table*}
In table \ref{tab:asd} is an overview of the results of all the simulations, for all the models studied, considering a constant mass loss, given by the value of $f$, which represents the fraction of mass that remains after a collision. The table shows the value of the final mass for the most massive object at the end of the simulation, and the percentage that the mass represents in comparison to the final mass when no mass loss is considered. All the values for mass on the table are averages.


\bsp	
\label{lastpage}
\end{document}